\documentclass{article}
\pdfoutput 1
\usepackage{graphicx}
\usepackage{amsmath}
\usepackage{amssymb}
\usepackage{pifont}% http://ctan.org/pkg/pifont
\newcounter{aux}

\DeclareFontFamily{OT1}{pzc}{}
\DeclareFontShape{OT1}{pzc}{m}{it}{<-> s * [1.10] pzcmi7t}{}
\DeclareMathAlphabet{\mathpzc}{OT1}{pzc}{m}{it}

\renewcommand{\tilde}{\widetilde}
\renewcommand{\bar}{\overline}

\title{Probing $|V_{td}|$ in single-top production at $pe^{\mp}$ colliders 
}

\author{Edier Paredes Cruz, Antonio O.\ Bouzas and F.\ Larios\\\small
  Departamento de F\'{\i}sica Aplicada, CINVESTAV-IPN \\\small
  Carretera Antigua a Progreso Km.\ 6, Apdo.\ Postal 73
  ``Cordemex''\\\small M\'erida 97310, Yucat\'an, M\'exico}
\date{}

\begin{document}

\maketitle

\begin{abstract}
  We study the consequences for top-quark physics of having electron
  and positron beams available at the LHeC and FCC-he, as was the case
  in HERA. We show that the asymmetry between top production in $pe^+$
  collisions and antitop production in $pe^-$ reactions is sensitive
  to $|V_{td}|$. By means of detailed parton-level Monte Carlo 
  simulations of single $t$ and $\bar{t}$ production and its
  backgrounds, we parametrize the asymmetry dependence on $|V_{td}|$
  and estimate its uncertainties.
  Our analysis includes realistic phase-space cuts, and
  machine-learning binary classifiers for background rejection.
  We thus obtain limits on $|V_{td}|$
  that are substantially stronger than current ones, and also smaller
  than current projections for the HL-LHC. We have
  $|V_{td}| < 1.6\times |V_{td}^\mathrm{PDG}|$ at the LHeC, at 68\%
  C.L.\ with $L_\mathrm{int}=2$/ab.
\end{abstract}

\section{Introduction}
\label{sec:intro}

The Cabibbo-Kobayashi-Maskawa (CKM) matrix elements are free
parameters in the Standard Model (SM), controlling the mixing among
down-type quarks in the charged-current electroweak interactions. They
are currently determined from low-energy physics processes. For the
third row of the CKM matrix, containing the elements that involve the
top quark, those are indirect determinations from experimental
measurements of $B^0$--$\bar{B}{}^0$ oscillations and kaon and
$B$-meson decays. A global fit to those data \cite{PDG}, assuming the
validity of the SM with three quark generations and unitarity of the
CKM matrix, results, in particular, in the absolute values of the
third row,
\begin{equation}
    \label{eq:ckm.pdg}
    |V_{td}^\mathrm{PDG}|=(8.57^{+0.20}_{-0.18})\times10^{-3},
    \quad
    |V_{ts}^\mathrm{PDG}|=(4.11^{+0.083}_{-0.072})\times10^{-2},
    \quad
    |V_{tb}^\mathrm{PDG}|=0.999118^{+3.1\times10^{-5}}_{-3.6\times10^{-5}}.
\end{equation}
In what follows we will take these values as reference.  The CMS
Collaboration has made a recent determination of third-row CKM matrix
elements, from measurements of single-top $t$-channel production and
decay in leptonic mode at the LHC at $\sqrt{s}=13$ TeV, yielding
$|V_{tb}|=0.988\pm0.024$ and $|V_{td}|^2+|V_{ts}|^2=0.06\pm0.06$
\cite{cmsckm20}. Notice that (\ref{eq:ckm.pdg}) is consistent with
these CMS values.

In recent years there has been an increasing interest in direct and
model independent measurements of the CKM mixing parameters at hadron
colliders. 
%Several measurement strategies have been put forth in the literature
%to determine third-row CKM elements at hadron colliders.
Measurements of %${\cal B}(t\to bW)$ and
the inclusive \cite{lac12}, or differential \cite{agu11}, cross
sections for $t$-channel single-top and $tW$ production were proposed
to set direct limits on $|V_{tq}|$.  Since a direct measurement of the
flavor off-diagonal top decay modes might be a way to measure
$|V_{td}|$ and $|V_{ts}|$, some strategies based on light-quark
tagging in $t\to Wq$ decay have also been proposed \cite{jan22,far22}.
In \cite{kam18}~the charge asymmetry in $tW$ associate production at
the LHC due to $d$, unlike $\bar{d}$, being a valence quark in the
proton, is proposed as a means to determine $|V_{td}|$. The
sensitivity to $|V_{td}|$ reported in \cite{kam18}, however, is rather
low, due to a combination of sizeable backgrounds, especially
$t\bar{t}$, and a weak dependence on $|V_{td}|$ of the asymmetry.  On
the other hand, a recent global analysis of $t$-channel single-top and
$tW$ associate production, based on published data on inclusive and
differential cross sections from the Tevatron CDF and D0, and the LHC
ATLAS and CMS experiments was carried out in \cite{cle19}. The global
fit of \cite{cle19} yields the absolute values of third-row CKM
matrix~elements:
%\\\noindent\fbox{\parbox{\textwidth}{
\begin{equation}
\label{eq:ckm.cle19}
\begin{array}{cccr}
        |V_{td}|=0.0^{+0.038}_{-0.000},&|V_{ts}|=0.0^{+0.069}_{-0.000},&|V_{tb}|=0.980^{+0.009}_{-0.012},&\mathrm{(marginal),}\\[2pt]
        |V_{td}|=0.0^{+0.023}_{-0.000},&|V_{ts}|=0.0^{+0.041}_{-0.000},&|V_{tb}|=0.986^{+0.008}_{-0.008},&\mathrm{(individual),}\\        
\end{array}
\end{equation}
%}}
where ``marginal'' refers to $|V_{tk}|$, $k=d,$ $s$, $b$ varied
independently, and ``individual'' to $|V_{td}|$ or $|V_{ts}|$ varied
one at a time, with $|V_{ts}|$ or $|V_{td}|=0$ and $|V_{tb}|=1$.  In
what follows we always set the CKM third row to their values
(\ref{eq:ckm.pdg}) and use the marginal value (\ref{eq:ckm.cle19}) as
a baseline for our results.

Electroweak top-quark production can also be probed in proton-lepton
collisions. The only such collider ever operated was HERA
\cite{bru22,wil21}, which provided collisions of a proton beam with
polarized electron and positron beams. HERA's center-of-mass energy of
320 GeV and total integrated luminosity of 800/pb were sufficient to
produce high-precision data on QCD and electroweak physics but not
enough to observe top production.  The latter will be measured at
future proton-lepton colliders,
%\\\fbox{\parbox{\textwidth}{
such as the Large Hadron-electron
Collider (LHeC) \cite{LHeC2020} (with beam energies $E_e=60$ GeV,
$E_p=7$ TeV, $\sqrt{s}=1.3$ TeV), and the Future Circular Collider
(FCC-he) \cite{FCC2019} (with $E_e=60$ GeV, $E_p=50$ TeV,
$\sqrt{s}=3.5$ TeV). For these $pe$ colliders the study of the
top-quark couplings to the Higgs and electroweak bosons will be among
the most important areas of research
\cite{LHeC2020,FCC2019,bou13,sar15,mel15,bor23}.
%}}
The dominant 
top-quark production mode at the LHeC and FCC-he is single-top
production, followed by top pair and associate $tW$ production
\cite{bou13}.

%\noindent{}\fbox{\parbox{\textwidth}{ \indent{}
HERA delivered
integrated luminosities in positron runs that were comparable to
those for electrons and, in its first stage, they were actually
somewhat larger, see figure 4 and table 1 in \cite{sou16}. A
different situation is considered in the baseline design for the
LHeC since, with current technology, the intensity of high-energy
positron beams would be smaller than that of electron ones by a
factor of 1/10--1/100 \cite{LHeC2012,LHeC2020}. There is, however,
great interest at present in the development of intense positron
beams for future $e^+e^-$ colliders such as CLIC \cite{CLIC19},
the proposed muon collider \cite{acc23}, and the LHeC and FCC-he
themselves. For the LHeC, new positron sources have been proposed
\cite{LHeC2020}: one \cite{zim18} would be based on photon beams
from the future LHC Gamma Factory \cite{kra15}, and another
\cite{zim19} based on an LHeC-generated free-electron laser
beam. Those sources would provide positron beams as intense as the
electron ones. For the FCC-he, which is decades in the future, it
is reasonable to assume that intense positron beams will be
available when it is operating.
%See \cite{LHeC2020}, p. 270 "Approaches towards LHeC Positrons"
%}}

%\noindent{}\fbox{\parbox{\textwidth}{\indent{}
In this paper we study the consequences for top-quark physics of    
having both electron and positron beams available at the LHeC and
FCC-he with similar luminosities, as was the case in HERA. For
simplicity, in what follows we assume the same luminosity for $e^\pm$
beams.
%}}
We show that the asymmetry between top production in $pe^+$
collisions and antitop production in $pe^-$ reactions, due to the
different distribution of $d$ and $\bar{d}$ in the proton, is
sensitive to $|V_{td}|$. By means of detailed parton-level Monte Carlo
simulations, we parametrize the asymmetry dependence on $|V_{td}|$ and
compute its statistical uncertainty. Furthermore, we carry out an
in-depth study of background processes and their charge asymmetries,
which contribute to the systematic uncertainty of the total
asymmetry. We thus obtain limits on $|V_{td}|$ that are substantially
stronger than current ones and smaller than current projections for
the HL-LHC \cite{cle19}.

This paper is organized as follows. In section \ref{sec:hdrn} we study
single-top production and decay in hadronic mode in $pe^\pm$
collisions. We discuss the signal and background processes and their
contributions to the charge asymmetry, and estimate the uncertainties
in the latter, to obtain limits on $|V_{td}|$. In section
\ref{sec:lptn}, we carry out a similar analysis for the leptonic mode
of single-top production. In section \ref{sec:finrem} we give our
summary and final remarks. Three appendices gather results used in the
previous sections. In appendix \ref{sec:polar} we discuss the impact
of beam polarization on the asymmetry and its sensitivity to
$|V_{td}|$. In appendix \ref{sec:app.asym} we derive the expression of
the asymmetry for the total process in terms of those for the signal
and backgrounds. In appendix \ref{sec:app.asym.stat} we give an
accurate estimate of the statistical uncertainty of the asymmetry in
terms of those of cross sections.

\section{Asymmetry in hadronic mode}
\label{sec:hdrn}

\begin{figure}
  \centering{}
 \includegraphics[scale=0.975]{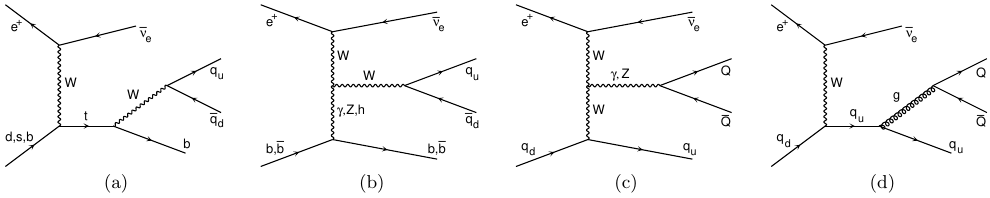}
  \caption{(a) Feynman diagrams for single-top production in $pe^+$
    collisions, in hadronic mode with flavor-diagonal decay. (b)
    Representative single-$W$ production diagram for the irreducible
    background $B_{1b}^+$. (c) and (d) Representative diagrams for the
    reducible background $B_{0b}^+$ (with $Q=q_u$, $q_d$ in (c) and
    $q_u$, $q_d$, $g$ in (d)), or
    $B_{2b}^+$ (with $Q=b$). In all cases, $q_u$, $q_d$ are as in
    (\ref{eq:signal}).
%    \fbox{
    The diagrams were drawn with \texttt{MadGraph5\_aMC@NLO}~\cite{alw14}.
%   }
    }
  \label{fig:dgrms}  
\end{figure}
We consider in this section single-top production in $pe^\pm$
collisions, and its decay in hadronic mode,
\begin{equation}
  \label{eq:signal}
  S^+: \;
  p e^+ \rightarrow t \bar{\nu}_e \rightarrow b\,q_u \bar{q}_d \bar{\nu}_e,
  \quad
  S^-: \;
  p e^- \rightarrow \bar{t} \nu_e \rightarrow \bar{b}\,\bar{q}_u q_d\nu_e,
  \qquad
  q_u=u, c,
  \quad
  q_d=d, s.  
\end{equation}
The processes $S^\pm$ are our signal processes. We decompose them into
subprocesses $S^\pm_d$ and $S^\pm_{sb}=S^\pm_s+S^\pm_b$, according to
the initial quark flavor. The Feynman diagrams for $S^+$ in the SM are
displayed in figure \ref{fig:dgrms} (a). There are a total of 12
diagrams for $S^+$, all of them $\mathcal{O}(\alpha^2)$, which include
all possible quark-flavor combinations with Cabibbo mixing only, and
as many for $S^-$. The scattering amplitudes for $S^\pm_d$ are
proportional to $V_{td}$, and those for $S^\pm_{s}$, $S^\pm_{b}$ to
$V_{ts}$ and $V_{tb}\sim1$, respectively. Therefore, the cross section
for $S^\pm$ is dominated by that for $S^\pm_{sb}$,
$\sigma_{S^\pm} = \sigma_{S^\pm_{sb}} + \sigma_{S^\pm_d} \simeq
\sigma_{S^\pm_{b}} $. Because $d$ is a valence quark in the proton,
however, the cross section for $S^+_d$ is substantially larger than
that for $S^-_d$, $S^\pm_{s}$, although still much smaller than that
for $S^\pm_{b}$ due to CKM suppression.

The parton level processes $S_{q_d}^+$ and $S_{q_d}^-$ ($q_d=d$, $s$,
$b$) are charge conjugate of each other.\footnote{
%  \fbox{\parbox{\textwidth}{
      More specifically,
  the initial states of $S_{q_d}^\pm$ are charge conjugate of each
  other in the unpolarized case. In what follows, we will always assume
  unpolarized beams except in appendix \ref{sec:polar}
%  }}
} We define the
charge asymmetry for a charge-conjugate pair of processes $P^\pm$ in
$pe^\pm$ collisions, such as $S_{q_d}^\pm$, as
\begin{equation}
  \label{eq:asym}
  A_P = \frac{\sigma_{P^+}-\sigma_{P^-}}{\sigma_{P^+}+\sigma_{P^-}}~,
\end{equation}
where $\sigma_{P\pm}$ is the cross section for process $P^{\pm}$,
which can be the signal process represented by the diagram (a) in
figure \ref{fig:dgrms}, or one of the backgrounds described by
diagrams (b)--(d) in the figure.  With the parton-level cross sections
computed with mild phase-space cuts
\begin{equation}
  \label{eq:mild.cuts}
p_T>4.5\,\mathrm{GeV}, \qquad |y| < 5.3,   
\end{equation}
for all quarks and gluons in the final state, the asymmetries for the
signal processes and its subprocesses are
$A_{S_d}\sim \mathcal{O}(10^{-1})$, and $A_{S_{sb}}$,
$A_S\sim \mathcal{O}(10^{-4})$.  Because the cross sections
$\sigma_{S^\pm}$ receive a large contribution from
$\sigma_{S^\pm_b} \gg \sigma_{S^\pm_d}$, their dependence on
$|V_{td,ts}|$ is very weak.  On the other hand, the contributions of
$S^\pm_{sb}$ to the numerator of the asymmetry $A_{S}$ largely cancel,
%\\\fbox{\parbox{\textwidth}{
leading to a strong dependence of $A_{S}$
on $|V_{td}|$, a more moderate one on $|V_{tb}|$ through the
denominator in (\ref{eq:asym}), and a negligible dependence on
$|V_{ts}|$. Since the current \emph{direct} limits on $|V_{tb}|$
bound it to within $\pm1\%$, as shown in (\ref{eq:ckm.cle19}), the
variation of $A_S$ on that narrow range of numerical values for
$|V_{tb}|$ is not significant. Therefore, in what follows we will
fix the value of $|V_{tb}|$ to that given in (\ref{eq:ckm.pdg}).
%  }}
Our goal in this paper is to determine what level of
sensitivity to $|V_{td}|$ can be reached through an experimental
measurement of $A_S$.

The irreducible backgrounds $B^\pm_{1b}$ to (\ref{eq:signal}) are
given by the same initial and final states as $S^\pm$, but with $bqq$
not originating in a top decay. They involve 774 diagrams each, out of
which 674 are purely electroweak $\mathcal{O}(\alpha^2)$, such as the
single-$W$ production diagram shown in Figure \ref{fig:dgrms} (b), and
the remaining 100 contain two QCD vertices and are
$\mathcal{O}(\alpha\alpha_s)$. The cross sections for $B_{1b}^\pm$, computed with
the lax phase-space cuts (\ref{eq:mild.cuts}), is a few percent of that for
$S_b$. It is dominated by diagrams with a $b$ (anti)quark in the
initial state since diagrams with a lighter initial parton are
suppressed by one power of $V_{ub,cb}$ (or $V_{td,ts}$ in diagrams
with a $t$-channel top). Because of this, as we discuss more
quantitatively below, the contribution to the asymmetry from this
process is small compared to that of the signal process. This
background is, therefore, easily controlled.

There are two types of reducible backgrounds. The first one consists
of processes of the form
\begin{equation}
  \label{eq:bck.0b}
  pe^+ \rightarrow j_1j_2j_3\bar{\nu}_e,
  \qquad
  pe^- \rightarrow j_1j_2j_3\nu_e,  
\end{equation}
with $j_n$ light partons. This includes top-production processes with
a $b$-quark in the initial state and the top decaying to $jW$. For
those processes, the cross section is suppressed by one power of
$V_{td,ts}$ relative to $S_b$, as with $S_d$, and their asymmetry is
as small as that of $S_{sb}$. Thus, as shown in Appendix
\ref{sec:app.asym}, their contribution to the total asymmetry is
negligibly small. We will not consider those processes further
here.\footnote{\label{fn1}However, we have taken these backgrounds
  fully into account in the computations of cross sections and
  asymmetries described below, thus explicitly verifying their
  smallness.} We denote $B^\pm_{0b}$ the reducible background
processes (\ref{eq:bck.0b}) not involving top production. Those are
given by 1440 Feynman diagrams each, out of which 1184 have
electroweak vertices only ($\mathcal{O}(\alpha^2)$) and 256 have two
QCD vertices ($\mathcal{O}(\alpha\alpha_s)$). Two representative
diagrams for $B^+_{0b}$ are shown in Figure \ref{fig:dgrms} (c) with
$Q=q_u$, $q_d$, and (d) with $Q=q_u$, $q_d$, $g$. The cross section
for $B_{0b}^\pm$, computed with the loose parton-level cuts
(\ref{eq:mild.cuts}), is about six times larger than that of
$S$. Furthermore, due to the subprocesses of (\ref{eq:bck.0b}) with
initial valence quarks, $B_{0b}$ has a large asymmetry
$A_{B_{0b}}$. These features make $B_{0b}^\pm$ the most important,
difficult to control backgrounds.

\addtocounter{footnote}{-1}
The second type of reducible background is given by processes of the
form (\ref{eq:bck.0b}) with $j_{1,2}=b$ or $\bar{b}$, and $j_3$ a
lighter parton. This includes top-production processes with a
$b$-quark in the initial state and the decay chain
$t\to bW \to b\bar{b}q_u$ or its charge conjugate. Similarly to the
previous case $B_{0b}$, those processes have both small cross section
and small asymmetry, so we will not consider them
further.\footnotemark\ We denote $B^\pm_{2b}$ the reducible background
processes (\ref{eq:bck.0b}) with two $b$ quarks in the final state not
involving top production. Those are given by 188 Feynman diagrams
each, out of which 172 have electroweak vertices only, and 16 have two
QCD vertices. Their nominal perturbative orders are as for $B_{0b}$. Two
representative diagrams for $B^+_{2b}$ are shown in Figure
\ref{fig:dgrms} (c), (d) with $Q=b$. The cross section for $B_{2b}^\pm$,
computed with the parton-level phase-space cuts described above, is
about 20 times smaller than that of $S$ at the LHeC energy, and about
50 times smaller at the FCC-he. However, $B_{2b}$ has a large
asymmetry $A_{B_{2b}}$, due to the valence quarks in the initial state
of $B^+_{2b}$, which makes it the second most important background to
the signal $S$ (\ref{eq:signal}).

%\noindent{}\fbox{\parbox{\textwidth}{

The other top production processes in a $pe$ collider are top-pair
production, $g e\rightarrow t\bar{t}e$, and associate $tW$
production, $b e\rightarrow tWe$. Both are neutral current (NC)
processes, with a cross section an order of magnitude smaller
\cite{bou13} than the single-top production (\ref{eq:signal}), and
a vanishing asymmetry (\ref{eq:asym}) at tree level. Furthermore,
due to their very different final states compared to
(\ref{eq:signal}), they should not contribute to the signal cross
section or asymmetry.
Another NC process worth mentioning is single $W$ production in
association with a $b$ quark, which contains the subprocess
$u e^+\rightarrow W^+be^+\rightarrow q_u\bar{q}_d b e^+$ having an
asymmetry $A$, as in (\ref{eq:asym}), proportional to $|V_{ub}|$.
Other NC processes with $b$-quark or gluon in the initial state, such
as $be^+\rightarrow b g g e^+$ or $g e^+\rightarrow g b \bar{b} e^+$,
are symmetric at tree level.
All this NC processes are rejected with perfect efficiency by the cuts
(\ref{eq:cuts}) at the parton level, due to the absence of missing
transverse energy, and with the inclusion of a veto on additional
charged leptons in the final state.
%}}   

We compute the tree-level cross section for single-top production and
its backgrounds with the matrix-element Monte Carlo generator
\texttt{MadGraph5\_aMC@NLO} (henceforth MG5) version 2.6.3
\cite{alw14}.  We use the parton distribution function (PDF)
%\fbox{
  \texttt{CTEQ66} from the library LHAPDF6 \cite{lhapdf6}.
%}
We
carry out the event analysis with \texttt{ROOT} version 6.22
\cite{root}, including the Toolkit for Multivariate Analysis,
\texttt{TMVA} \cite{TMVA2007}.  We apply in our MG5 simulations the
mild, flavor-blind cuts (\ref{eq:mild.cuts}) to avoid infrared
divergences in background processes.  We perform event pre-selection
in our ROOT analysis. We $b$-tag partons in each event by assigning to
each parton a tag $\tau_b(J)=0$ if $J$ is a light parton, and 1 if it
is a $b$ quark. Here, $J_{0,1,2}$ are the final state partons
(``jets'') ordered by decreasing $p_T$. We use a working point with a
$b$-tagging efficiency $\eta_b=0.85$, and mistagging probabilities
$p_c=0.1$ for $c$ quarks and $p_j=0.01$ for lighter partons.  We then
apply the following phase-space cuts on $b$-tagged events,
\begin{equation}
\label{eq:cuts}
\begin{gathered}
   |y(J)| < 5.0, \quad |y(J_b)|<3.0,\\
   p_T(J) > 5.0\,\mathrm{GeV}, \quad
   \Delta R(J,J') > 0.4, \quad
   \not\!E_T>10.0\,\mathrm{GeV},\\
   142.0 < m(J_0,J_1,J_2) < 202.0\,\mathrm{GeV}.
\end{gathered}
\end{equation}
On the first line of this equation we require all partons ($J$) to be
inside the detector, and $b$-tagged ones ($J_b$) in the central region where
$b$-tagging is operational and efficient. On the second line of
(\ref{eq:cuts}) we impose minimal $p_T$ and isolation requirements on
all final-state particles. On the third line we require that the three
final-state partons reconstruct the top mass.

%\noindent{}\fbox{\parbox{\textwidth}{
The asymmetry $A$ in (\ref{eq:asym}) depends non-linearly on the
signal and background cross sections. As discussed in detail in
appendix \ref{sec:app.asym}, because of this non-linearity, the
measured asymmetry $A$ is given in terms of the signal and backgrounds
asymmetries $A_S$, $A_{B_0}$, \ldots, as a sum of weighted asymmetries
$\tilde{A}_S$, $\tilde{A}_{B_0}$, \ldots, with weights that are ratios
of cross sections. The exact expression for these asymmetries
$\tilde{A}$ is given in equation (\ref{eq:eq8}). In Table
\ref{tab:tab1} we show the cross sections and the associated
asymmetries $A$ from (\ref{eq:asym}) and $\widetilde{A}$ from
(\ref{eq:eq8}), for the signal (\ref{eq:signal}), the irreducible
background $B_{1b}$ and the reducible backgrounds $B_{0b}$, $B_{2b}$,
at LHeC and FCC-he energies, with the cuts (\ref{eq:cuts}). The large
asymmetries of the reducible background processes can be seen in
that table.
%}}
\newcommand{\marg}[1]{\rotatebox{90}{\makebox[-7.5pt]{#1}}}
\begin{table}
  \centering{}
  \begin{tabular}{c|c|cccc|cccc}
    \multicolumn{2}{c}{} & \multicolumn{4}{c}{LHeC}& \multicolumn{4}{|c}{FCC-he}\\\cline{3-10}
\multicolumn{1}{c}{cuts}&\multicolumn{1}{c}{}&$S$&$B_{1b}$&$B_{0b}$&$B_{2b}$&$S$&$B_{1b}$&$B_{0b}$&$B_{2b}$\\\hline
    &$\sigma_{P^+}$ [fb] &952.261     &17.933      &4117.950    &30.243       &7826.525     &116.790     &22609.2       &103.833 \\\cline{2-10}
    \marg{(\ref{eq:cuts})}
    &$\sigma_{P^-}$ [fb] &950.786      &17.934      &5458.860    &50.152       &7829.969     &116.402     &23752.2       &116.472 \\\cline{2-10}
    &$A\times10^4$\rule{0.0pt}{10pt}
                         &7.752&-0.223&-1400&-2476&-2.200&16.64&-246.5&-573.7 \\\cline{2-10}
    &$\widetilde{A}\times10^4$\rule{0.0pt}{12pt}
                         &1.272&-0.0006899&-1156&-17.27&-0.5513&0.06211&-183&-2.023 \\\hline\hline        
    &$\sigma_{P^+}$ [fb] &115.625&0.110&0.827&0.0142&1253.402&1.0409&4.438&0.0434\\\cline{2-10}
    \marg{(\ref{eq:cuts})+(\ref{eq:score.cuts})}
    &$\sigma_{P^-}$ [fb] &114.831&0.111&0.795&0.0285&1253.620&1.067&4.715&0.0468\\\cline{2-10}
    &$A\times10^4$\rule{0.0pt}{10pt}
                         &34.44&-37.44&191.7&-3351&-0.8698&-125.6&-302.8&-372.8\\\cline{2-10}
    &$\widetilde{A}\times10^4$\rule{0.0pt}{12pt}
                         &34.16&-0.03564&1.343&-0.6166&-0.8658&-0.1051&-1.101&-0.01335\\\hline            
  \end{tabular}
  \caption{Cross sections in fb and their associated asymmetries $A$,
    (\ref{eq:asym}), and $\widetilde{A}$, (\ref{eq:eq8}), at LHeC and
    FCC-he energies. The results of applying both the cuts
    (\ref{eq:cuts}) and those plus (\ref{eq:score.cuts}) are
    shown. Asymmetries are computed from cross sections before
    rounding the latter to three decimal digits at most.}
  \label{tab:tab1}  
\end{table}

The signal processes $S$, (\ref{eq:signal}), and the background
processes $B_{kb}$, $k=0$, 1, 2, have different phase-space
distributions, which we can use to reject backgrounds.  We resort to
standard machine-learning methods such as boosted-decision trees
(BDTs) (see, e.g., \cite{coa22} and refs.\ therein) that are known to
be effective tools to optimize background rejection.  For instance,
BDTs were applied in LHC analyses of Higgs production and decay
\cite{cms12,cms14}.  We use gradient-boosted BDTs (BDTGs), implemented
in the TMVA \cite{TMVA2007} framework with $10^3$ trees and default
parameters. In all cases we use a set of $10^6$ events passing the
cuts (\ref{eq:cuts}) for training, and a separate, independent set of
$5\times10^5$ to $10^6$ events for application. Furthermore, we apply the
trained BDTGs also on the combined training-plus-application data for
control.

\begin{figure}
  \includegraphics[scale=0.9]{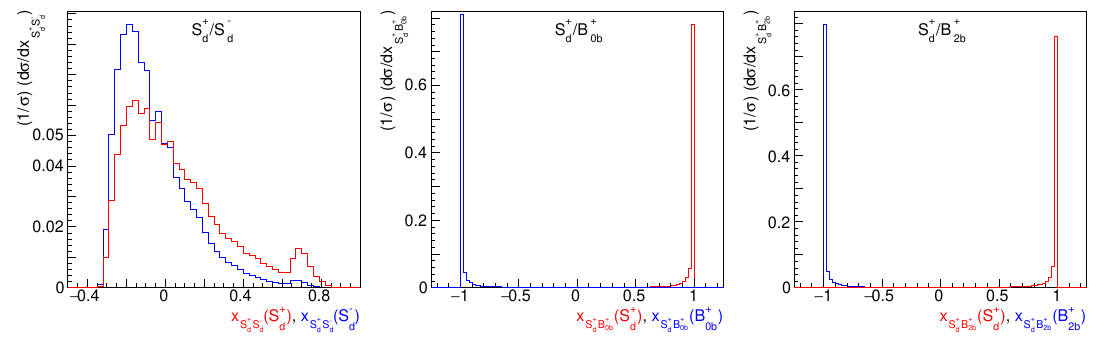}
  \caption{Normalized differential cross sections with respect to the
    score variable for the signal (red line) and background (blue
    line),
    %\fbox{
      at the LHeC energy,
    %}
    for three BDTs trained with the
    signal-and-background pairs indicated.}
  \label{fig:fgr_c}
\end{figure}

We use three different classifiers: first, we train a BDTG with the
process $S_d^+$ as signal and $B_{0b}^+$ as background, denoted
BDT$(S_d^+/B_{0b}^+)$, to suppress $B_{0b}^\pm$ with respect to the
signal. Second, similarly, we train a BDT$(S_d^+/B_{2b}^+)$ to reject
the backgrounds $B_{2b}^\pm$. Third, we also train a BDT with $S_d^+$
as signal and $S_d^-$ as background, BDT$(S_d^+/S_d^-)$, to enhance
the signal asymmetry.  Each BDT is trained with 28 kinematic variables,
containing the rapidities $y$, transverse momenta $p_T$, azimuthal
angles $\varphi$, and invariant masses of one or more partons. We
consider also the $p_T$ and $\varphi$ of the missing transverse
energy, corresponding to the neutrino in (\ref{eq:signal}). We also
include the $b$-tags $\tau_b$ defined above, the number $N_b$ of $b$
quarks, the phase-space distances among partons
$\Delta R=\sqrt{(\Delta y)^2+(\Delta\varphi)^2}$, and the cosine
$\cos(\theta_{Wj}^*)$ of the angle between the 3-momenta of the
reconstructed $W$ and its decay product with the largest $p_T$ in the
rest frame of the reconstructed $W$ (analogous to the leptonic
variable defined in \cite{agu13}, see next section). Furthermore, we
consider the cosine $\cos(\theta_{j\mathrm{spct}})$ of the angle
between the $W$ decay-product with largest $p_T$ and the parton not
coming from $W$ decay, or ``spectator,'' in the lab frame, the cosine
$\cos_{(\theta_{J_{0}J_{1}})}$ of the angle between the three-momenta
of $J_{0,1}$ in lab frame, the sphericity $S_\mathrm{chrgd}$ and
aplanarity $A_\mathrm{chrgd}$ for charged particles, as defined in
chapter 9 of \cite{barger}, and also the ``visible''
$S_\mathrm{vsbl}$, $A_\mathrm{vsbl}$, computed with the four-momenta
of the charged particles and the missing four-momentum
$\not\!p^\mu=(|\!\not\!\vec{p}_T|,\not\!p^x_T,\not\!p^y_T,0)$. The 20
variables highest ranked in BDTG training for each signal and
background combination are as follows:
\begin{equation}
    \label{eq:variables}
\begin{aligned}
  S_d^+/S_d^-:& \; y_{(J_0,J_1,J_2)}, \varphi_{(J_1)}, p_{T (\not
    E_T)}, \varphi_{(J_2)}, \varphi_{(J_0,J_1,J_2)}, y_{(J_2)}, \Delta
  R_{(J_0,J_2)}, \varphi_{(\not E_T)}, m_{(J_1,J_2)},
  \cos_{(\theta_{Wj}^*)}, \varphi_{(J_0)}, \\
  &\; S_\mathrm{chrg}, \Delta R_{(J_0,J_1)},
  p_{T(J_2)}, \Delta R_{(J_1,J_2)}, m_{(J_0,J_1)}, y_{(J_1)},
  \cos_{(\theta_{J_{0}J_{1}})}, p_{T(J_{1})}, m_{(J_0,J_2)}, 
  \\
  S_d^+/B_{0b}^+:&\; m_{(J_0,J_1,J_2)}, m_{(J_1,J_2)},
  m_{(J_0,J_2)}, m_{(J_0,J_1)}, p_{T (\not E_T)}, y_{(J_0,J_1,J_2)},
  p_{T(J_2)}, y_{(J_0)}, y_{(J_2)}, p_{T(J_0)}, \cos_{(\theta_{Wj}^*)},\\
  &\; \tau_{b(J_1)}, p_{T(J_1)}, \cos_{(\theta_{Wj}^*)}, \tau_{b(J_2)},
  y_{(J_1)}, \tau_{b(J_0)}, N_b, \Delta R_{(J_1,J_2)},
  S_\mathrm{vsbl},
  \\
  S_d^+/B_{2b}^+:& \; m_{(J_0,J_1)}, m_{(J_0,J_2)}, m_{(J_1,J_2)},
  m_{(J_0,J_1,J_2)}, \tau_{b(J_0)}, \tau_{b(J_1)}, p_{T (\not E_T)},
  y_{(J_2)}, \cos_{(\theta_{Wj}^*)}, y_{(J_0,J_1,J_2)}, N_b, \\
  &\; y_{(J_0)}, \tau_{b(J_2)}, y_{(J_1)}, A_\mathrm{chrgd},
  p_{T(J_0)}, p_{T(J_2)}, \cos_{(\theta_{\mathrm{jspct}})}, p_{T(J_1)}, 
  \Delta R_{(J_0,J_2)}.
\end{aligned}    
\end{equation}
In the first line we see that the most discriminating variable is the
top rapidity, which is clearly related to the process $S^+_d$ being
initiated by a valence quark, unlike $S^-_d$. Similarly, on the second
and third lines of (\ref{eq:variables}), the highest ranked variables
are masses and $b$-tags, as expected for reducible backgrounds less
resonant than the signal. In figure \ref{fig:fgr_c} we display the
differential cross section
% \fbox{
  at the LHeC energy
% }
with respect to
the score variable for the signal and background processes for the
three BDTs trained as in (\ref{eq:variables}).
The corresponding results for the FCC-he are similar to those for the
LHeC in figure \ref{fig:fgr_c}. Those distributions motivate us to
apply the following selection cuts on our event samples to reject the
backgrounds,
\begin{equation}
  \label{eq:score.cuts}
  x_{S_d^+S_d^-}>0.0,
  \quad
  x_{S_d^+B_{0b}^+}>0.99,
  \quad
  x_{S_d^+B_{2b}^+}>0.96,    
\end{equation}
where $x_{P_1P_2}$ refers to the score variable generated by
BDT($P_1/P_2$) trained with process $P_1$ as signal and $P_2$ as
background, as described above. 
The cross sections and asymmetries for the signal and background
processes computed with the preselection cuts (\ref{eq:cuts}) and the
selection ones (\ref{eq:score.cuts}) are shown in Table
\ref{tab:tab1}. We notice, from the table, the high degree of
background rejection, which, in turn, leads to the signal asymmetries
$A_S$ and $\widetilde{A}_S$ being approximately equal, and the
background ones $\widetilde{A}_B$ ($B=B_{kb}$, $k=0,$ 1, 2) being
smaller than that of the signal, especially at the LHeC.
%In what
%follows we take those background contributions
%$\widetilde{A}_{B_{kb}}$, added in quadrature, as estimates of the
%systematic uncertainty of the asymmetry $\widetilde{A}_S$. That
%uncertainty, in turn, added in quadrature to the statistical one
%(\ref{eq:six}), yields the total uncertainty of the asymmetry
%$\Delta_\mathrm{tot}\widetilde{A}_S$.

We stress here that the cuts on score variables (\ref{eq:score.cuts})
play a role completely analogous to phase-space cuts in cut-based
analyses such as \cite{kam18}. In fact, the cuts (\ref{eq:score.cuts})
entail cuts on the phase-space variables (\ref{eq:variables}) used to
compute the score variables. Thus, using a BDT provides optimized
phase-space cuts on the relevant kinematic variables.  In particular,
the cut on $x_{S_d^+S_d^-}$ selects the phase-space region where the
asymmetry is largest. Notice that in \cite{kam18}, section III B is
dedicated to finding an optimal set of cuts to maximize the charge
asymmetry defined in that reference. By using the cut on
$x_{S_d^+S_d^-}$ in (\ref{eq:score.cuts}), we enable the BDT to find
the optimal region for us.

%\noindent{}\fbox{\parbox{\textwidth}{
In order to estimate the total
uncertainty in the asymmetry $A$, we take into account the
background contributions to it, $\widetilde{A}_{B}$, the
statistical uncertainties as described in appendix
\ref{sec:app.asym.stat}, eq.\ (\ref{eq:six}), and the systematic
uncertainties related to the choice of PDF and its scale. The
latter are provided by MG5 and the library LHAPDF6 for the cross
sections, and propagated to the asymmetry as in appendix
\ref{sec:app.asym.stat}. We obtain uncertainties of 4.5\% (scale)
and 3.34\% (PDF) at the LHeC, and 23.5\% (scale) and 35.7\% (PDF)
at the FCC-he. The relative asymmetry uncertainties at the FCC-he
are significantly larger than those at the LHeC due to the
smallness of the asymmetry in the former collider relative to the
latter. Obtaining an estimate of systematic uncertainties related
to finite energy resolution and other detector effects is beyond
the scope of this paper. However, since the uncertainty in the
asymmetry is dominated by the statistical one, we expect those
other systematic uncertainties to have a moderate impact on the
total uncertainty. We obtain our estimate of the total asymmetry
uncertainty %$\Delta_\mathrm{tot}\widetilde{A}_S$
$\Delta_{\widetilde{A}_S,\mathrm{tot}}$ by adding in
quadrature the three types of uncertainty mentioned above:
statistical, background and systematic.
%}}

\begin{figure}
  \centering{}  
  \includegraphics[scale=0.45]{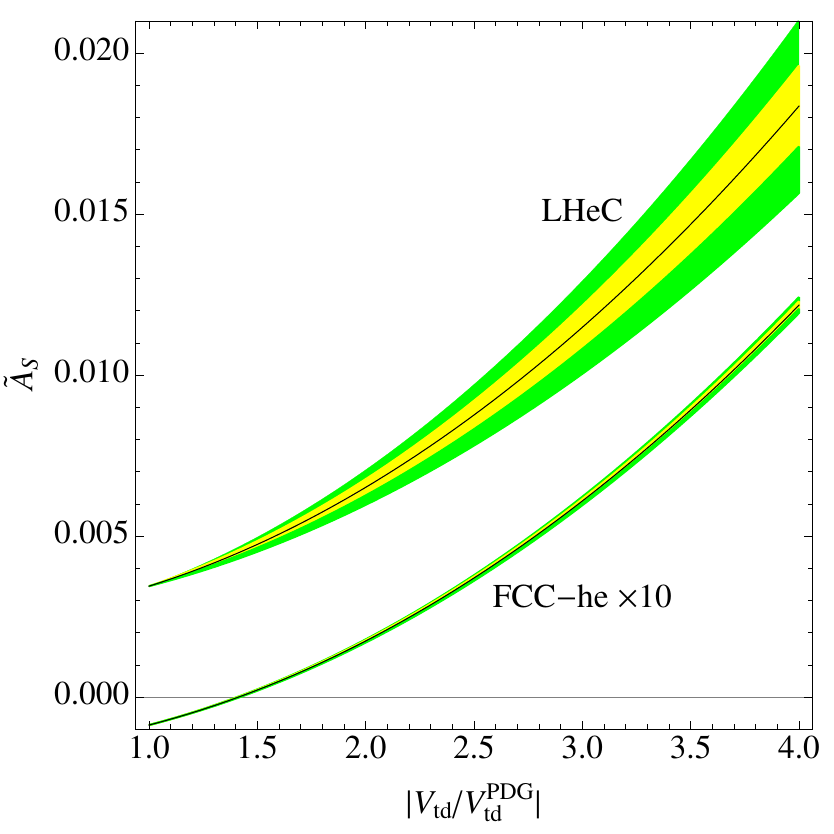}
  \caption{The polynomial fits (\ref{eq:rat0}) with the parameters
    given in the text, at the LHeC and FCC-he energies, with 68\%
    (yellow) and 95\% (green) CL bands.}
  \label{fig:fits}
\end{figure}

To obtain bounds on $|V_{td}|$, we consider the dependence on it of
the signal asymmetry $\widetilde{A}_S(|V_{td}|)$.  For a given total
uncertainty, the allowed values for $|V_{td}|$ are determined with
statistical significance $\mathcal{S}$ ($\mathcal{S}=1,$ 2, \ldots) by
the inequality
\begin{equation}
  \label{eq:ineq}
  \widetilde{A}_S(|V_{td}|) < \widetilde{A}_S^\mathrm{SM} +
  \mathcal{S}%\Delta_\mathrm{tot}\widetilde{A}_S,
  \Delta_{\widetilde{A}_S,\mathrm{tot}},   
\end{equation}
where $\tilde{A}_S^\mathrm{SM} = \tilde{A}_S(|V_{td}^\mathrm{PDG}|)$
and $|V_{td}^\mathrm{PDG}|$ from (\ref{eq:ckm.pdg}).  We parametrize
$\widetilde{A}_S(|V_{td}|)$ as,
\begin{equation}
  \label{eq:rat0}
    \tilde{A}_S(|V_{td}|) = \tilde{A}_S^\mathrm{SM} + a
      \left(\frac{|V_{td}|}{|V_{td}^\mathrm{PDG}|}-1\right)
      + b \left(\frac{|V_{td}|}{|V_{td}^\mathrm{PDG}|}-1\right)^2 ,
\end{equation}
% \fbox{
with $|V_{td}^\mathrm{PDG}|$ from (\ref{eq:ckm.pdg}).
% }
We
determine the parameters in (\ref{eq:rat0}) from an extensive set of
Monte Carlo simulations at each energy, to which (\ref{eq:rat0}) is
fitted. We obtain excellent fits with $a=2.117\times10^{-3}$,
$b=9.511\times10^{-4}$ at the LHeC, and $a=1.185\times10^{-4}$,
$b=5.858\times10^{-5}$ at the FCC-he.
% \\\fbox{
These parameter values, as well
as their 68\% and 95\% CL intervals shown in figure \ref{fig:fits},
were computed with Mathematica \cite{wolf}.
%}
Notice that these values imply,
in particular, that the asymmetry $\tilde{A}_S(|V_{td}|)$ is an
increasing function of $|V_{td}|$. This is very different from the
asymmetry defined in \cite{kam18} in the context of $pp$ collisions at
the LHC, which is a \emph{decreasing} function of $|V_{td}|$, as shown
in figure 5 of \cite{kam18}.

% \fbox{
Substituting (\ref{eq:rat0}) into (\ref{eq:ineq}) yields an upper
bound on $|V_{td}|$.
%}
\begin{table}
  \centering{}
  \begin{tabular}{c|cc|cc|cc|cc}
    &\multicolumn{4}{c}{LHeC}&\multicolumn{4}{|c}{FCC-he}\\\hline
C.L. &\multicolumn{2}{c}{68\%}&\multicolumn{2}{c}{95\%}&\multicolumn{2}{|c}{68\%}&\multicolumn{2}{c}{95\%} \\\hline
$L_\mathrm{int}$ [1/ab] & 1 & 2 & 1 & 2 & 1 & 2 & 1 & 2\\\hline
\rule{0pt}{11pt}    $|V_{td}|/|V_{td}^\mathrm{PDG}|<$ & 1.80 & 1.61 & 2.34 & 2.04 & 2.99 & 2.59 & 4.10 & 3.52\\
  \end{tabular}
  \label{tab:tab2}
  \caption{Limits on $|V_{td}|$ obtained from simulated
    charge-asymmetry measurements in single-top production in hadronic
    mode in $pe^{\pm}$ collisions. Results are shown for two values of
    Confidence Level corresponding to $\mathcal{S}=1$, 2, and two
    values of integrated luminosity.
%    \fbox{
      The value of
      $|V_{td}^\mathrm{PDG}|$ is given in (\ref{eq:ckm.pdg}).
%    }
  }
\end{table}
These bounds are the main results of this paper. They are reported in
Table \ref{tab:tab2}, at statistical significance $\mathcal{S}=1,$ 2,
for two values of integrated luminosity at the energies of the LHeC
and FCC-he. We see from the table that the limits obtained at the LHeC
energy are stricter than those at the FCC-he, as expected, since due
to the evolution of the proton PDF with the energy scale, the
asymmetry is expected to be smaller at higher energies, and its
dependence on $|V_{td}|$ correspondingly weaker.

The best current direct bounds on $|V_{tk}|$, $k=d$, $s$, $b$, are
obtained in \cite{cle19} from LHC data, as shown above in
(\ref{eq:ckm.cle19}).  With the notation of table \ref{tab:tab2} and
(\ref{eq:ckm.pdg}), those values and the projection for the HL-LHC
\cite{cle19}, are written as,
  \begin{equation}
    \label{eq:cle19}
    \frac{|V_{td}|}{|V_{td}^\mathrm{PDG}|} < \left\{
    \begin{array}{cr}
      4.43& \text{(LHC marginal)}\\
      2.68& \text{(LHC individual)}\\
      2.89& \text{(HL-LHC)}      
    \end{array}\right.
 \qquad (68\% \mathrm{C.L.}).
\end{equation}
%\fbox{\parbox{\textwidth}{
We stress the large difference between the marginal and individual LHC
limits in this equation, arising from the global analysis in
\cite{cle19} being based on observables having a strong dependence on
$|V_{tb}|$, but a poor one on $|V_{td,ts}|$. On the other hand, as
discussed in the text under (\ref{eq:mild.cuts}), the asymmetry
(\ref{eq:asym}) is sensitive to $|V_{td}|$, but has little variation
over the interval of $|V_{tb}|$ determined by the direct bounds
(\ref{eq:ckm.cle19}) from \cite{cle19}. Thus, in a global analysis of
third-row CKM parameters including the asymmetry $A$, there should be
a small difference between marginal and individual limits on
$|V_{td}|$. In this sense, the 68\% CL limits on $|V_{td}|$ reported
in table \ref{tab:tab2} should be compared to the marginal limits in
(\ref{eq:cle19}).
%}}
We notice, however, that the results in table \ref{tab:tab2} at the
LHeC energy, at 68\% CL, are substantially stronger than even the
individual limit in (\ref{eq:cle19}).

\section{Leptonic mode}
\label{sec:lptn}

In leptonic mode the signal processes take the form
\begin{equation}
  \label{eq:signal.lpt}
  S^+: \;
  p e^+ \rightarrow t \bar{\nu}_e \rightarrow b\,\ell^+ \nu_\ell \bar{\nu}_e,
  \quad
  S^-: \;
  p e^- \rightarrow \bar{t} \nu_e \rightarrow \bar{b}\,\ell^- \bar{\nu}_\ell\nu_e,
  \qquad
  \ell = e, \mu,
\end{equation}
to be compared to (\ref{eq:signal}). The asymmetry of the signal is
smaller in leptonic mode than in hadronic mode: with the cuts
(\ref{eq:cuts}), $A_S$ is about one-third of its hadronic mode value.
The leptonic-mode analogs of the irreducible backgrounds $B_{1b}^\pm$ and
the reducible one $B_{0b}^\pm$ are obtained, similarly to the signal
(\ref{eq:signal.lpt}), by replacing a pair of light partons with a
lepton-neutrino pair. In this mode there are no background processes
analog to $B_{2b}^\pm$, with two $b$ quarks in the final state.
Background processes have smaller cross sections relative to signal
than in hadronic mode. With the cuts (\ref{eq:cuts}), the irreducible
background $B_{1b}^\pm$ cross section remains at about 2\% that of the
signal, as in hadronic mode. On the other hand, the reducible
backgrounds $B_{0b}^\pm$ cross sections are 36\% that of the signal in this
mode, compared to six times larger than the signal in hadronic mode.

Because there are two neutrinos in the final state, their
three-momenta cannot be easily reconstructed, so we use all kinematic
variables analogous to those in (\ref{eq:variables}) related to the
final-state charged particles, $b$ and $\ell$. However, we also
approximately reconstruct the three-momentum of the neutrino produced
in $t$ decay, $\mathrm{\nu_\mathrm{pr}}$, by minimizing the function
\begin{equation}
    \label{eq:chi2}
    \chi^2 = \frac{1}{\Gamma_t^2}
    \left(m(J,\ell,\nu_\mathrm{pr})\right) - m_t)^2 +
    \frac{1}{\Gamma_W^2}
    \left(m(\ell,\nu_\mathrm{pr})\right) - m_W)^2
\end{equation}
with respect to $\vec{p}(\nu_\mathrm{pr})$, with the mass and decay
width parameters from \cite{PDG}. With this three-momentum, we can
reconstruct the top and $W$ momenta, and the transverse momentum of
the scattered neutrino $\nu_\mathrm{sc}$ as missing transverse
momentum.

Similarly to the hadronic mode discussed in the previous section, we
train two BDTs: one, BDT$(S_d^+/B_{0b}^+)$, to suppress $B_{0b}^\pm$
with respect to the signal; the other, BDT$(S_d^+/S_d^-)$, to enhance
the signal asymmetry. We use 25 variables to train the BDTs. The 20
highest ranked for each signal and background combination are as
follows:
\begin{equation}
    \label{eq:variables.lpt}
\begin{aligned}
  S_d^+/S_d^-:& \; m_{(J,\ell)}, y_{(\ell)}, y_{(J,\ell)}, y_{(J)},
  y_{(J,\ell,\nu_\mathrm{pr,rc})}, \cos_{(\theta_{W\ell}^*)},
  p_{T(J)}, p_{T(\ell)}, \varphi_{(J)}, m_{T(W_\mathrm{rc})},
  \\
  &\;  \varphi_{(\ell)}, E_{(J)}, P_\mathrm{vsbl},
  \Delta\varphi_{(\nu_\mathrm{pr,rc}, \nu_\mathrm{sc,rc})}, 
  dR_{(J,\ell)}, m_{T(t_\mathrm{rc})}, S_{T\mathrm{chrg}},
  S_{T\mathrm{vsbl}}, S_\mathrm{chrg}, \varphi_{(J,\ell)},
  \\
  S_d^+/B_{0b}^+:&\; p_{T(\ell)}, p_{T(J)}, m_{(J,\ell)},
  p_{T(J,\ell)}, y_{(J)}, y_{(\ell)}, N_b, y_{(t_\mathrm{rc})},
  E_{(\ell)}, \Delta R_{(J,\ell)}, \cos_{(\theta_{W\ell}^*)},
  \\
  &\; m_{T(t_\mathrm{rc})},
  \Delta\varphi_{(\nu_\mathrm{pr,rc}, \nu_\mathrm{sc,rc})}, E_{(J)},
  P_\mathrm{vsbl}, S_{T\mathrm{chrg}}, m_{T(W_\mathrm{rc})},
  \cos_{(\theta_{J\ell})}, S_{T\mathrm{vsbl}}, A_{\mathrm{vsbl}}. 
  \end{aligned}    
\end{equation}
The variable $\theta_{W\ell}^*$ is the angle between the directions of
the charged-lepton and the $W$ 3-momenta in the $W$ rest frame
\cite{agu13}. The sphericities and aplanarities
$S_{\mathrm{chrgd},\mathrm{vsbl}}$,
$A_{\mathrm{chrgd},\mathrm{vsbl}}$, for charged particles and
``visible'' ones, are defined as in the previous section above
(\ref{eq:variables}). In the leptonic case we also include the
planarity $P=(2/3)(S-2A)$ and the transverse sphericity $S_{T}$
\cite{sas21}, both charged and visible. Because the background cross
sections are smaller, softer cuts than in hadronic mode
(\ref{eq:score.cuts}) can be used to suppress them,
\begin{equation}
  \label{eq:score.cuts.lpt}
  x_{S_d^+S_d^-}>0.0,
  \quad
  x_{S_d^+B_{0b}^+}>0.9,
\end{equation}
We have applied the cuts on score variables for BDTs trained with the
25 variables described above, and also for BDTs trained without
reconstructed variables. Comparison of the two methods shows that the
inclusion of reconstructed variables improves background rejection by
20\%.

The dependence of the asymmetry $\widetilde{A}_S$ on $|V_{td}|$ is
quadratic, and we parametrize it as in (\ref{eq:rat0}). In leptonic
mode that dependence is weaker than in the hadronic one, with
$a=1.471\times10^{-3}$, $b=7.016\times10^{-4}$ at the LHeC and
$a=1.098\times10^{-4}$, $b=5.413\times10^{-5}$ at the FCC-he. As in
the hadronic mode, the parameters for the LHeC are roughly 15 times
larger than those for the FCC-he. On the other hand, the parameters
$a$, $b$ in hadronic mode are approximately 40\% larger than those in
leptonic mode at the LHeC, and about 10\% larger at the FCC-he.  That
weaker dependence results in a reduced sensitivity to $|V_{td}|$ in
leptonic mode, as reflected in the limits shown in table
\ref{tab:tab3}, which are weaker than those from hadronic mode in
table \ref{tab:tab2}. We notice, however, that the 68\% C.L.\ limits
in table \ref{tab:tab3} are all smaller than the marginal limit
in (\ref{eq:cle19}).
\begin{table}
  \centering{}
  \begin{tabular}{c|cc|cc|cc|cc}
    &\multicolumn{4}{c}{LHeC}&\multicolumn{4}{|c}{FCC-he}\\\hline
C.L. &\multicolumn{2}{c}{68\%}&\multicolumn{2}{c}{95\%}&\multicolumn{2}{|c}{68\%}&\multicolumn{2}{c}{95\%} \\\hline
$L_\mathrm{int}$ [1/ab] & 1 & 2 & 1 & 2 & 1 & 2 & 1 & 2\\\hline
\rule{0pt}{11pt}    $|V_{td}|/|V_{td}^\mathrm{PDG}|<$ &2.28 & 2.00&3.06&2.63&4.32&3.68&6.03&5.10\\
  \end{tabular}
  \label{tab:tab3}
  \caption{Limits on $|V_{td}|$ obtained from simulated
    charge-asymmetry measurements in single-top production in leptonic
    mode in $pe^{\pm}$ collisions. Results are shown for two values of
    Confidence Level corresponding to $\mathcal{S}=1$, 2, and two
    values of integrated luminosity.}
\end{table}

\section{Final remarks}
\label{sec:finrem}

\begin{figure}
  \centering{}  
  \includegraphics[scale=0.6]{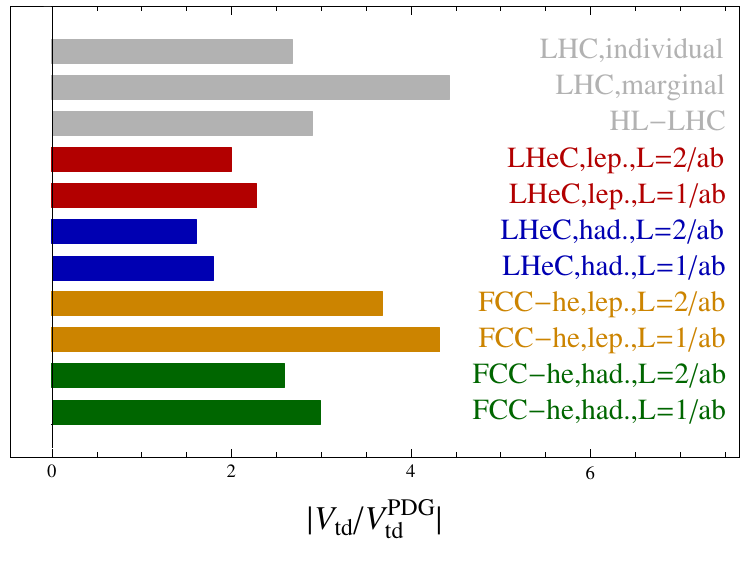}
  \caption{Limits on $|V_{td}|$ at 68\% CL from single-top production
    at the LHeC and FCC-he in hadronic and leptonic mode, see tables
    \ref{tab:tab2}, \ref{tab:tab3}, and from \cite{cle19}, see eq.\ (\ref{eq:cle19}).}
  \label{fig:barc}
\end{figure}
     
In this work we considered the possibility that the LHeC and the
FCC-he will run not only in $pe^-$ but also in $pe^+$ mode,
%\fbox{
with similar luminosities in both modes,
% }
as HERA did.  In that
case, the charge asymmetry in single-top production in $pe^\mp$
collisions is sensitive to $|V_{td}|$.

In the hadronic mode of single-top production, BDTs provide sufficient
background suppression to obtain bounds on $|V_{td}|$, at the parton
level, that are substantially more restrictive than current
experimental ones.  As reported in table \ref{tab:tab2}, at the LHeC
energy we obtain $|V_{td}|/|V_{td}^\mathrm{PDG}|<1.80$ at 68\%
C.L. with $L_\mathrm{int}=1/$ab, and $<1.6$ with
$L_\mathrm{int}=2/$ab, which are significantly better than the current
LHC limits (\ref{eq:cle19}) of 4.43 (marginal).  Our limits are also
smaller than those projected for the HL-LHC in table 4 of
\cite{cle19}, $|V_{td}|/|V_{td}^\mathrm{PDG}|<2.89$ (68\% CL). At the
FCC-he the asymmetry in single-top production is smaller than at the
LHeC, and has a weaker dependence on $|V_{td}|$, so that the bounds
obtained are weaker than at the LHeC, though still stronger than the
marginal ones in (\ref{eq:cle19}), as also shown in table
\ref{tab:tab2}.  In leptonic mode the dependence of the asymmetry
$\widetilde{A}_S$ on $|V_{td}|$ is weaker than in hadronic mode. This
leads to looser limits on $|V_{td}|$,
%\fbox{
as shown in table \ref{tab:tab3},
% }
even though the backgrounds in that mode are
smaller than in the hadronic one. Still the limits obtained at 68\%
C.L.\ at the LHeC are stricter than the current ones in
(\ref{eq:cle19}) from \cite{cle19}.
% \fbox{
The 68\% CL limits on $|V_{td}|$
from tables \ref{tab:tab2} and \ref{tab:tab3}, and from \cite{cle19}
as shown in (\ref{eq:cle19}), are summarized in figure
\ref{fig:barc}.
%}

We expect to carry out a more detailed analysis including QCD
evolution and fast detector simulation. More importantly, by including
other observables in the analysis besides the charge asymmetry studied
here, the sensitivity to $|V_{td}|$ of single-top production in
$pe^\mp$ collisions could be improved, both at the LHeC and especially
at the FCC-he. 
%\\\fbox{\parbox{\textwidth}{
In addition, we will address the possibility of potential
contributions that may appear at the NLO level.  We do not expect
them to change the conclusions reached here because, in contrast to
the LHC study made in [7], at the LHeC both signal and background
contributions to the asymmetry appear already at LO.
%}}
We will report those results elsewhere.

We remark, finally, that because the $pe^\pm$ asymmetry in single-top
production has a different dependence on third-generation CKM matrix
elements from that of top observables in hadron colliders, due to the
distinct underlying physics, its measurement would make a very
important contribution to future global analyses comprising data from
the LHC, HL-LHC, LHeC, future $e^+e^-$ and/or $\mu^+\mu^-$ colliders,
and the FCC-hh and FCC-he.

\paragraph*{Acknowledgements}

We thank Georgina Espinoza Gurriz for assistance with computer
hardware. This work was partially supported by Sistema Nacional
de Investigadores de Mexico.

\paragraph*{Data Availability Statement}

No data associated in the manuscript.

\appendix{}

\section{
%  \fbox{
The effect of beam polarization
%  }
}
\label{sec:polar}

In the foregoing we always assumed unpolarized beams. However, future
$pe$ colliders are expected to have the capability to polarize lepton
beams longitudinally, with polarizations reaching up to 80\%. In this
appendix we briefly discuss the effects of beam polarization on the
asymmetry and its sensitivity to $|V_{td}|$.

For charged-current scattering, such as the signal processes
(\ref{eq:signal}), the cross-section dependence on longitudinal beam
polarization is linear,
\begin{equation}
    \label{eq:polar.1}
    \sigma_{S^\pm}^\mathrm{pol} = (1\pm\mathcal{P}_{e^\pm})
    \sigma_{S^\pm},
\end{equation}
with $-1\leq\mathcal{P}_{e^\pm}\leq1$ the beam polarization and
$\sigma_{S^\pm}$ the cross section for the unpolarized
process. Whereas in the unpolarized case the cross-section asymmetry
(\ref{eq:asym}) is also a charge asymmetry, in the polarized case we
have to distinguish between the two. We define the cross-section
asymmetry by analogy to (\ref{eq:asym}) as,
  \begin{equation}
    \label{eq:asym.polar}
    A^{\mathrm{pol}}_{S,\sigma} =
    \frac{\sigma_{S^+}^\mathrm{pol}-\sigma_{S^-}^\mathrm{pol}}{\sigma_{S^+}^\mathrm{pol}+\sigma_{S^-}^\mathrm{pol}}~. 
  \end{equation}
It will be convenient in what follows to define the quantities,
\newcommand{\polave}{\ensuremath{\overline{\mathcal{P}}}}
\newcommand{\dpol}{\ensuremath{\Delta\mathcal{P}}}
\begin{equation}
  \label{eq:polave}
  \polave = \frac{1}{2}(\mathcal{P}_{e^+}+\mathcal{P}_{e^-}),
  \qquad
  \dpol = \frac{1}{2}(\mathcal{P}_{e^+}-\mathcal{P}_{e^-}).  
\end{equation}
We have $-1\leq\polave,\dpol\leq1$. Notice, however, that $\polave$
and $\dpol$ are not independent, since they must satisfy
$|\mathcal{P}_{e^\pm}| = |\polave\pm\dpol|\leq1$. From the definitions
(\ref{eq:asym}), (\ref{eq:polar.1}), (\ref{eq:asym.polar}),
(\ref{eq:polave}), we obtain the identity,
\begin{equation}
  \label{eq:polar.id}
  A^{\mathrm{pol}}_{S,\sigma} =
  \frac{\polave+(1+\dpol)A}{1+\dpol+\polave A},
\end{equation}
where $A$ is the unpolarized asymmetry.  Equation (\ref{eq:polar.id})
shows that, as expected, beam polarization can induce a cross-section
asymmetry even if $A=0$. Such polarization asymmetry is independent of
the interactions underlying the processes $S^\pm$. Equation
(\ref{eq:polar.id}) can be inverted, which motivates the following
definition of charge asymmetry for polarized beams,
\begin{equation}
  \label{eq:charge.polar}
      A^{\mathrm{pol}}_{S,\mathrm{ch}} =   \frac{(1+\dpol)
        A^{\mathrm{pol}}_{S,\sigma} -\polave}{1+\dpol-\polave
        A^{\mathrm{pol}}_{S,\sigma}}, 
\end{equation}
with $A^{\mathrm{pol}}_{S,\sigma}$ from (\ref{eq:asym.polar}). If
$\polave=0$ we have
$\mathcal{P}_{e^+}=-\mathcal{P}_{e^-}\equiv\mathcal{P}$,
$\dpol=\mathcal{P}$, and
$A^{\mathrm{pol}}_{S,\mathrm{ch}}=A^{\mathrm{pol}}_{S,\sigma}=A$, so
that the charge asymmetry is independent of $\mathcal{P}$.  In this case,
$\polave=0$, of which the unpolarized beams considered in the previous
sections are a particular case, the effect of beam polarization on the
sensitivity of the asymmetry to $|V_{td}|$ can be readily quantified.
Because the measurement uncertainty for the asymmetry is dominated by
the statistical one, the cross-section dependence on $\mathcal{P}$
will affect that sensitivity through the corresponding effect on the
number of events. By using (\ref{eq:polar.1}), we can write
$N^\mathrm{pol} = \sigma_{S^\pm}^\mathrm{pol} L_\mathrm{int} =
\sigma_{S^\pm} L_\mathrm{int}^\mathrm{eff},$ with the ``effective''
integrated luminosity
$L_\mathrm{int}^\mathrm{eff} = (1+\mathcal{P}) L_\mathrm{int}$. Thus,
for a run with $L_\mathrm{int}=1/$ab, in the limit case
$\mathcal{P}=+1$ we have $L_\mathrm{int}^\mathrm{eff}=2/$ab and we can
read the limits on $|V_{td}|$ from table \ref{tab:tab2}. For other
values of $L_\mathrm{int}^\mathrm{eff}$, those limits can be obtained
from table \ref{tab:tab2} by interpolation of the two luminosity
values shown there.

\section{Asymmetry in signal and background processes}
\label{sec:app.asym}

Let $S$ be the signal process, and $B_k$, $1\leq k\leq N_B$,
background processes. We assume first that there is no interference
among those processes, as is the case for reducible backgrounds with
different number of $b$ quarks in the final state. We then have,
\begin{equation}
  \label{eq:eq6}
  \sigma^{\pm}_{S+B} = \sigma^{\pm}_S + \sum_{k=1}^{N_B} \sigma^{\pm}_{B_k}.
\end{equation}
For a given process $P$, signal or background, we have the charge
asymmetry (\ref{eq:asym}), with (\ref{eq:eq6}) leading to the
algebraic identity, 
\begin{equation}
  \label{eq:eq8}
  \begin{gathered}
    A_{S+B} = \widetilde{A}_S + \sum_{k=1}^{N_B} \widetilde{A}_{B_k}~,
    \qquad
    \widetilde{A}_S = \rho A_S~,
    \qquad
    \widetilde{A}_{B_k} = \rho \eta_k A_{B_k}~,\\
    \eta_k = \frac{\sigma^{+}_{B_k}+\sigma^{-}_{B_k}}{\sigma^{+}_{S}+\sigma^{-}_{S}}~,    
    \qquad
    \rho
    %=\left(
    %  1+\sum_{h=1}^{N_B}\frac{\sigma^{+}_{B_h}+\sigma^{-}_{B_h}}{\sigma^{+}_{S}+\sigma^{-}_{S}}
    %\right)^{-1}
    =\frac{1}{1+\sum_{k=1}^{N_B}\eta_k} ~,
  \end{gathered}    
\end{equation}
expressing the asymmetry of the total process $S+B$ in terms of the
asymmetries and cross sections of the individual processes $S$ and
$B_k$. We stress that the contributions to the total asymmetry
$A_{S+B}$ of the individual background asymmetries $A_{B_k}$ are
suppressed by the cross-section ratios $\eta_k$.  If there is
interference among the processes $S$, $B_k$, the expression
(\ref{eq:eq8}) holds in the approximation that such interference is
small and can be neglected. The validity of (\ref{eq:eq8}) in those
cases must be checked numerically.

\section{Approximate normal distribution for the asymmetry}
\label{sec:app.asym.stat}

Let us consider the cross sections $\sigma_\pm$ to be jointly normal
distributed,
\begin{equation}
  \label{eq:one}
  (\sigma_+,\sigma_-) \sim \mathcal{N} \exp\left(
    -\frac{1}{2(1-\rho^2)}\left(
      \frac{(\sigma_+-\mu_{\sigma_+})^2}{\Delta_{\sigma_+}^2}-2\rho
      \frac{(\sigma_+-\mu_{\sigma_+})(\sigma_--\mu_{\sigma_-})}{\Delta_{\sigma_+}\Delta_{\sigma_-}}
      +\frac{(\sigma_--\mu_{\sigma_-})^2}{\Delta_{\sigma_-}^2}      
    \right)
  \right)
\end{equation}
where $\mu_{\sigma_\pm}$ are the means and $\Delta_{\sigma_\pm}>0$ the
  standard deviations for $\sigma_{\pm}$, and $-1<\rho<1$ the
  correlation parameter. 
We assume that the probability for $\sigma_\pm$ to be negative is
exponentially small, which means that $0<\Delta_{\sigma_\pm} \ll
\mu_{\sigma_\pm}$. We assume also that the mean asymmetry is small:
$|(\mu_{\sigma_+}-\mu_{\sigma_-})/(\mu_{\sigma_+}+\mu_{\sigma_-})|
\lesssim 0.02$. Then, to a very good approximation, the asymmetry
$A=(\sigma_+-\sigma_-)/(\sigma_++\sigma_-)$
is normally distributed with
\begin{equation}
  \label{eq:four}
    \mu_A =
    \frac{\mu_{\sigma_+}-\mu_{\sigma_-}}{\mu_{\sigma_+}+\mu_{\sigma_-}}~,
    \qquad
    \Delta_A^2 = \frac{\Delta_{\sigma_+}^2+\Delta_{\sigma_-}^2-2\rho
      \Delta_{\sigma_+}\Delta_{\sigma_-}}
    {(\mu_{\sigma_+}+\mu_{\sigma_-})^2}~.
\end{equation}
In the particular case of the statistical fluctuations of a Poisson
process we have
$\Delta_{\sigma_\pm}/\mu_{\sigma_{\pm}} = 1/\sqrt{N_\pm}$, where
$N_\pm=\sigma_\pm L_\mathrm{int}$ is the number of events, assuming for
simplicity the same integrated luminosity $L_\mathrm{int}$ for $pe^+$
and $pe^-$ collisions. We then have from (\ref{eq:four}),
\begin{equation}
  \label{eq:six}
    \left(\Delta_A^2\right)_\mathrm{stat} = 
    \frac{1}{(N_++N_-)^2} \left(N_++N_- - 2\rho\sqrt{N_+N_-} \right)
    \simeq \frac{1-\rho}{N_++N_-}~,
\end{equation}
where in the last step we used the approximate equality
$N_+\simeq N_-$ due to the smallness of the asymmetry. Equation
(\ref{eq:six}) gives the statistical uncertainty for $A$.

\end{document}